\documentclass[pre,superscriptaddress,twocolumn,footinbib]{revtex4-1}
\pdfoutput=1
\usepackage[colorlinks=true,urlcolor=blue]{hyperref}
\usepackage{amsfonts}
\usepackage{graphicx}
\usepackage{placeins}
\usepackage{amsmath}
\usepackage{amssymb}
\usepackage{xcolor}
\usepackage{bm}
\usepackage{layouts}

\definecolor{red}{rgb}{0.75,0,0}
\definecolor{blue}{rgb}{0,0,0.75}
\definecolor{green}{rgb}{0,0.5,0}

\newcommand{\dd}{\mathrm{d}}

\begin{document}
	
\title{Chiral stresses in nematic cell monolayers}

\author{Ludwig A. Hoffmann}
\affiliation{Instituut-Lorentz, Leiden University, P.O. Box 9506, 2300 RA Leiden, The Netherlands}
\author{Koen Schakenraad}
\affiliation{Instituut-Lorentz, Leiden University, P.O. Box 9506, 2300 RA Leiden, The Netherlands}
\affiliation{Mathematical Institute, Leiden University, P.O. Box 9512, 2300 RA Leiden, The Netherlands}
\author{Roeland M. H. Merks}
\affiliation{Mathematical Institute, Leiden University, P.O. Box 9512, 2300 RA Leiden, The Netherlands}
\affiliation{Institute of Biology, Leiden University, P.O. Box 9505, 2300 RA Leiden, The Netherlands}
\author{Luca Giomi}
\thanks{Corresponding author: giomi@lorentz.leidenuniv.nl}
\affiliation{Instituut-Lorentz, Leiden University, P.O. Box 9506, 2300 RA Leiden, The Netherlands}

\begin{abstract}
Recent experiments on monolayers of spindle-like cells plated on adhesive stripe-shaped domains have provided a convincing demonstration that certain types of  collective phenomena in epithelia are well described by active nematic hydrodynamics. While recovering some of the hallmark predictions of this framework, however, these experiments have also revealed a number of unexpected features that could be ascribed to the existence of chirality over length scales larger than the typical size of a cell. In this article we elaborate on the microscopic origin of chiral stresses in nematic cell monolayers and investigate how chirality affects the motion of topological defects, as well as the collective motion in stripe-shaped domains. We find that chirality introduces a characteristic asymmetry in the collective cellular flow, from which the ratio between chiral and non-chiral active stresses can be inferred by particle-image-velocimetry measurements. Furthermore, we find that chirality changes the nature of the spontaneous flow transition under confinement and that, for specific anchoring conditions, the latter has the structure of an imperfect pitchfork bifurcation. 
\end{abstract}

\maketitle

\section{Introduction}
Multicellular systems of prokaryotes, such as suspensions of planktonic bacteria, have historically played a pivotal role in the development of the hydrodynamics of active fluids, since the early work of Batchelor on the stress distribution in suspensions of microswimmers \cite{Batchelor:1970}. By contrast, multicellular systems of eukaryotes have entered only recently into the realm of active hydrodynamics, following a number of inspiring experimental works on epithelial and mesenchymal cell layers and tissues (see e.g. Refs. \cite{Duclos:2014,Garcia:2015,Duclos:2016,Saw:2017,Kawaguchi:2017,Blanch-Mercader:2018,Duclos:2018}). Among these, monolayers of clotured spindle-like cells, such as NIH 3T3 mouse embryo fibroblasts \cite{Duclos:2016}, murine neural progenitor cells (NPCs) \cite{Kawaguchi:2017}, human bronchial epithelial cells (HBEC) \cite{Blanch-Mercader:2018}, Retinal Pigment Epithelial (RPE1) cells \cite{Duclos:2018} and C2C12 mouse myoblasts \cite{Duclos:2018} represent an especially promising class of model systems, because of their connection with active nematic liquid crystals (see e.g. Ref. \cite{Doostmohammadi:2018}). 

First identified as a broken symmetry in certain types of cell cultures \cite{Kemkemer:2000a}, and later exploited to decipher their static \cite{Kemkemer:2000b} and dynamical properties \cite{Duclos:2014,Duclos:2016,Saw:2017,Kawaguchi:2017,Blanch-Mercader:2018,Duclos:2018}, nematic order has surged as one of the central themes in collective cell dynamics. In layers of spindle-like cells, where the local orientation can be unambiguously identified, nematic order is marked by the presence of $\pm 1/2$ disclinations \cite{Duclos:2016,Kawaguchi:2017}, i.e. point-like singularities about which the average cellular orientation rotates by $\pm \pi$. Consistently with the predictions of active nematic hydrodynamics \cite{Giomi:2013,Giomi:2014}, these defects self-propel and pairwise annihilate until cell crowding freezes the system into a jammed configuration. Before dynamical arrest, the collective motion of the cells gives rise to a decaying turbulent flow at low Reynolds number, whose statistics, spatial organization and spectral structure are in exceptional agreement with the hydrodynamic picture \cite{Giomi:2015} (but see Ref. \cite{Henkes:2019} for an alternative theoretical picture based on glassy dynamics). 

Another remarkable demonstration of active hydrodynamic behavior in eukaryotic cell layers has been recently reported by Duclos {\em et al.}, upon confining spindle-like RPE1 and C2C12 cells within adhesive stripe-shaped domains \cite{Duclos:2018}. Depending on the width of the stripe the system was found either in a stationary state, with the cells parallel to the longitudinal direction of the confining region, or in a collectively flowing state characterized by a spontaneous tilt of the cells toward the center of the stripe. The latter picture, often referred to as spontaneous flow transition, had been anticipated for over a decade by Voituriez {\em et al.} \cite{Voituriez:2005} and represents one of the hallmarks of active liquid crystals. While confirming this seminal prediction, however, Duclos {\em et al.} have also highlighted a number of unexpected features that could be ascribed to the existence of chirality over length scales larger than the typical size of a cell.

The notion of chirality is not new in active matter and has been theoretically explored well before the interest around collective cell dynamics in eukaryotes had started to blossom. F\"urthauer {\em et al.}, for instance, demonstrated that microscopic torque dipoles, such as those arising from rotating molecular motors or flagella, give rise to antisymmetric stresses and angular momentum fluxes, which, in turn, drive rotating flows and other chiral patterns on the large scale \cite{Furthauer:2012a,Furthauer:2012b}. More recently, Banerjee {\em et al}. showed that rotational motion at the microscopic scale further enriches the spectrum of hydrodynamical behaviors or chiral active fluids by giving rise to non-dissipative ``odd'' viscosity \cite{Banerjee:2017}, analogous to that found in quantum Hall fluids \cite{Avron:1995}. Whereas undoubtedly interesting and relevant for a broad class of biological and synthetic systems, these mechanisms appear however unsuited to account for the chirality observed in the experiments by Duclos {\em et al.}, because of the manifest lack of rotational motion at the scale of individual cells.  

In this article, we show that macroscopic chirality can arise in nematic cell layers as a consequence of a misalignment between the cell's local orientation and active forces, even in the absence of microscopic rotational motion (Sec. \ref{sec:chiral_stresses}). Collectively, this gives rise to a chiral and yet symmetric stress tensor that, in two dimensions, complies with the symmetries of the nematic phase. Next, we explore the effect of such a chiral stress on the active flow generated by $\pm 1/2$ disclinations and identify a characteristic signature of chirality from which the ratio between chiral and non-chiral active stresses can be experimentally estimated (Sec. \ref{sec:defects}). Finally, following Duclos {\em et al.} \cite{Duclos:2018}, we investigate the hydrodynamic stability of a chiral nematic cell monolayer confined on adhesive stripes and subject to various boundary conditions and classify all possible scenarios arising from the interplay between the geometry of the confining region, the extensile/contractile stresses and chirality (Sec. \ref{sec:adhesive_stripes}).
  
\section{\label{sec:chiral_stresses}Chiral stresses}

\begin{figure}[t]
\centering
\includegraphics[width=0.97\columnwidth]{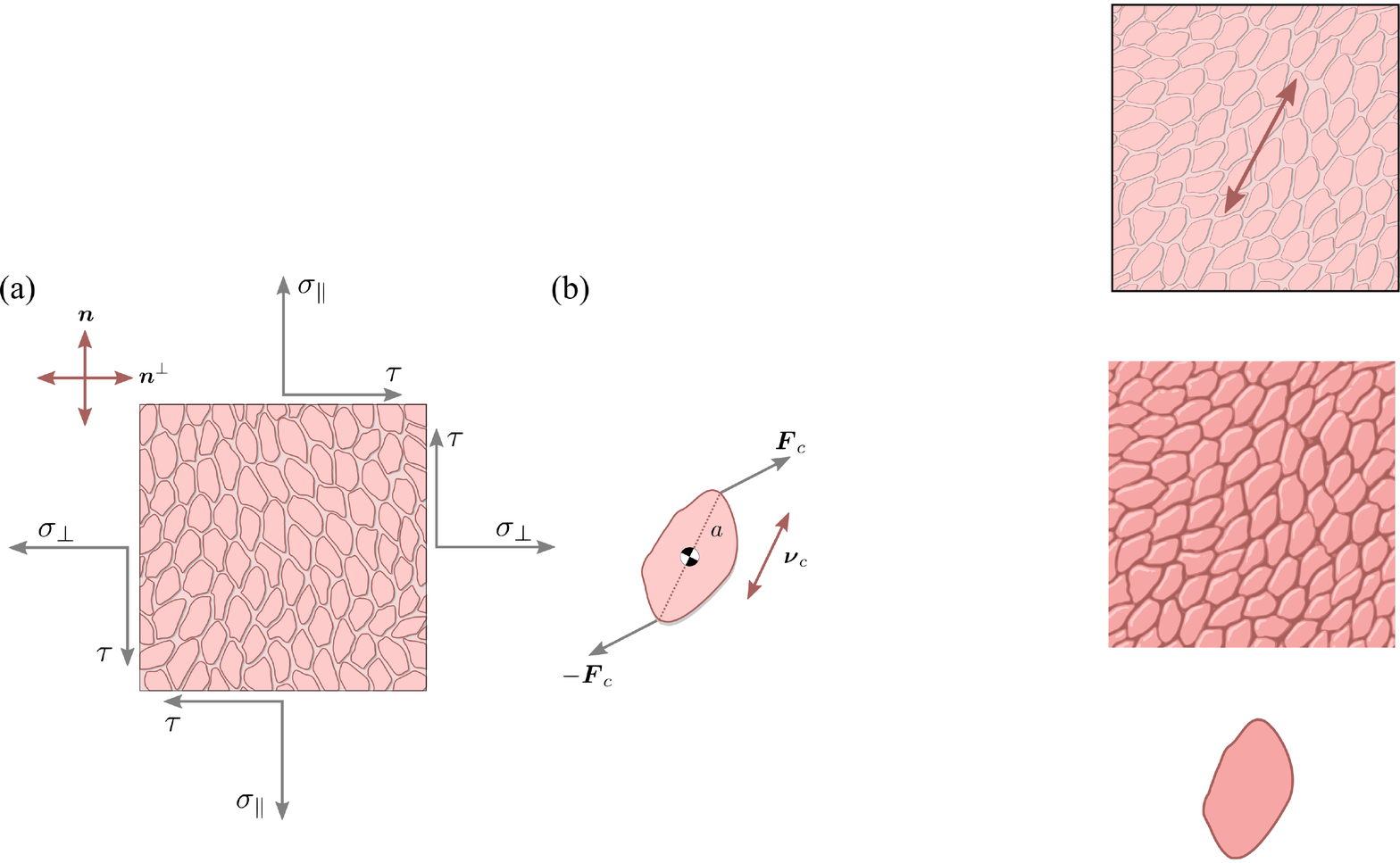}	
\caption{\label{fig:figure1}Schematic representation of stresses and forces in cellular nematic monolayers. (a) A volume element whose faces are conventionally oriented in the direction of the nematic director $\bm{n}$ and its normal vector $\bm{n}^{\perp}$. In the most general setting, the volume element is subject to three independent stresses: the two normal stresses $\sigma_{\parallel}$ and $\sigma_{\perp}$ and the shear stress $\tau$. (b) The chiral stress $\tau$ arises at the microscopic level when the force exerted by an individual cell is neither parallel nor perpendicular to the cell axis.}
\end{figure}

Let us consider a two-dimensional volume element and let $\bm{n}$ be the nematic director representing the average direction of the enclosed cells (Fig. \ref{fig:figure1}a). The most generic form of the stress tensor associated with such a volume element can be expressed in the basis of the nematic director $\bm{n}$ and its normal vector $\bm{n}^{\perp}$, namely:
\begin{equation}\label{eq:nematic_stress}
\bm{\sigma}^{\rm a} = \sigma_{\parallel}\bm{n}\bm{n}+\sigma_{\perp}\bm{n}^{\perp}\bm{n}^{\perp}+\tau(\bm{n}\bm{n}^{\perp}+\bm{n}^{\perp}\bm{n})\;.
\end{equation}
Here $\sigma_{\parallel}$ and $\sigma_{\perp}$ represent the stresses experienced by the volume element in the direction parallel and perpendicular to $\bm{n}$, whereas $\tau$ is the shear stress. By construction, $\bm{\sigma}^{\rm a}$ is invariant under the transformation $\bm{n}\rightarrow-\bm{n}$, thus is consistent with the symmetry of the nematic phase, but is evidently chiral for any non-zero $\tau$ value, as the stress distribution depicted in Fig. \ref{fig:figure1}a does not coincide with its mirror image. Furthermore, taking $\bm{n}^{\perp}\bm{n}^{\perp}=\boldsymbol{I}-\bm{n}\bm{n}$, with $\bm{I}$ the identity tensor, allows one to cast the active stress as the sum of an isotropic contribution, equivalent to an active pressure, a deviatoric term, common to all active nematic liquid crystals regardless of their chirality \cite{Pedley:1992,Simha:2002}, and a chiral term, namely:
\begin{equation}\label{eq:active_stress}  
\bm{\sigma}^{\rm a} 
= -P^{\rm a}\bm{I}+\alpha\left(\bm{n}\bm{n}-\frac{1}{2}\bm{I}\right)+\tau(\bm{n}\bm{n}^{\perp}+\bm{n}^{\perp}\bm{n})\;,
\end{equation}
where: 
\begin{equation}
P^{\rm a}=-\frac{\sigma_{\parallel}+\sigma_{\perp}}{2}\;,\qquad
\alpha=\sigma_{\parallel}-\sigma_{\perp}\;.
\end{equation}
Microscopically, the chiral stress $\tau$ might originate from the fact that the force exerted by an individual cell is tilted with respect to the cell orientation. To illustrate this concept let us consider an individual cell whose major and minor axes are parallel to the unit vectors $\bm{\nu}_{c}$ and $\bm{\nu}_{c}^{\perp}$ (Fig. \ref{fig:figure1}b). Following Lau and Lubensky \cite{Lau:2009}, one can express the stress tensor as  $\bm{\sigma}^\text{a} = \sum_{c} \bm{d}_{c}\delta(\bm{r}-\bm{r}_{c})$, where the index $c$ runs over all the cells in the system, $\bm{r}_{c}$ is the position of the $c-$th cell and
\begin{equation}\label{eq:force_dipole}
\bm{d}_{c} = \frac{1}{2} \oint_{\Sigma_{c}} \dd A \left(\bm{f}_{c} \bm{R}_{c} + \bm{R}_{c}\bm{f}_{c} \right) \;,
\end{equation}
is a force-dipole tensor. In Eq. \eqref{eq:force_dipole}, $\bm{R}_{c}$ is the distance between the cell's surface and center of mass, $-\dd A\,\bm{f}_{c}$ the force exerted by the cell's area element $\dd A$ on the surrounding medium and the integral is calculated over the surface $\Sigma_{c}$ of the $c-$th cell. A complete derivation can be found in Ref. \cite{Lau:2009}. For an effectively two-dimensional system, such as the one considered here, $\dd A = \dd \ell\,h$, with $h$ the thickness of the cell along the $z-$direction, here assumed to be uniform throughout the sample, and $\ell$ the one-dimensional arc-length. 

Now, the magnitude of the active stresses $P^{\rm a}$, $\alpha$, and $\tau$ depend exclusively upon the distribution of the forces exerted by the cells along their contour. The simplest approximation of the force density field $\bm{f}_{c}$ consists then of a dipole of the form:
\begin{equation}\label{eq:force_density}
\bm{f}_{c} = \bm{F}_{c}\delta(\bm{R}_{c}-a\bm{\nu}_{c})-\bm{F}_{c}\delta(\bm{R}_c+a\bm{\nu}_{c})\;,	
\end{equation}
where $a$ is the cell's major semi-axis (Fig. \ref{fig:figure1}b and Ref. \cite{Simha:2002}). To make progress, we express the cellular forces in the $\{\bm{\nu}_{c},\bm{\nu}_{c}^{\perp}\}$ basis, namely $\bm{F}_{c}=F_{\parallel}\bm{\nu}_{c}+F_{\perp}\bm{\nu}_{c}^{\perp}$, with $F_{\parallel}$ and $F_{\perp}$ the longitudinal and transverse components of the force exerted by the $c-$th cell, here assumed for simplicity uniform throughout the system. Replacing this in Eq. \eqref{eq:force_density} and coarse-graining $\bm{\sigma}^{\rm a}$ over the length scale of a volume element $\Omega(\bm{r})$ centered at $\bm{r}$ (Fig. \ref{fig:figure1}a) yields:
\begin{equation}
\bm{\sigma}^{\rm a} = a \rho \left[2F_{\parallel} \langle \bm{\nu}_{c}\bm{\nu}_{c} \rangle + F_{\perp} \langle \bm{\nu}_{c}\bm{\nu}_{c}^{\perp}+\bm{\nu}_{c}^{\perp}\bm{\nu}_{c}\rangle\right]\;,
\end{equation}
where $\langle \cdots \rangle$ is the spatial average within $\Omega(\bm{r})$ and $\rho$ is the local cell number density. Finally, calculating the averages and comparing with Eq. \eqref{eq:active_stress} readily gives:
\begin{equation}\label{eq:microscopic}
P^{\rm a} = - a\rho F_{\parallel}\;,\qquad
\alpha = 2 a\rho S F_{\parallel}\;,\qquad
\tau = a\rho S F_{\perp}\;, 
\end{equation}
where $S=2\langle |\bm{\nu}_{c}\cdot\bm{n}|^{2}\rangle-1$ is the local nematic order parameter in two dimensions. As expected, in the absence of nematic order (i.e. $S=0$), the active forces exerted by the cells result exclusively in an effective pressure, while both the deviatoric and chiral stress vanishes identically. When nematic order is not uniform throughout the system, hence $S$ varies in space, the active stress tensor of Eq. \eqref{eq:active_stress} is more conveniently expressed in terms of tensor order parameter $Q_{ij}=S(n_{i}n_{j}-\delta_{ij}/2)$, namely: 
\begin{equation}
\sigma_{ij}^{\rm a}=-P^{\rm a}\delta_{ij}+\alpha_{0}Q_{ij}-2\tau_{0}\epsilon_{ik}Q_{kj}\;,
\end{equation}
where $\alpha_{0}=2a\rho F_{\parallel}$ and $\tau_{0}=a\rho F_{\perp}$ are constants independent on the nematic order parameter $S$ and $\epsilon_{ij}$ is the two-dimensional Levi-Civita tensor (i.e. $\epsilon_{xx}=\epsilon_{yy}=0$ and $\epsilon_{xy}=-\epsilon_{yx}=1$).

Some comments are in order. Although Eq. \eqref{eq:force_density} is only a rudimental approximation of the force field generated by an irregularly-shaped cell, considering a more involved force distribution does not change the qualitative picture with respect to the emergence of chiral stresses, as long as this is asymmetric with respect to the cell's longitudinal direction. To illustrate this concept, we discuss in Appendix \ref{sec:appendix} the case of a quadrupolar force distribution. Whereas the exact origin of this asymmetry is beyond the scope of the present article, the biophysical literature is not scarce of examples where chirality can be detected at the single-cell level. For instance, various mammalian cells, when plated on micropatterns, can break the left-right symmetry by suitably positioning their internal organelles with respect to the cell body \cite{Wan:2011}. Analogously, chirality can emerge at the scale of the entire cell from the self-organization of the actin cytoskeleton \cite{Tee:2015, Schakenraad:2019}. The broken symmetry can furthermore propagate over the mesoscopic scale and bias the cell's collective migratory motion \cite{Worely:2015}.

\section{\label{sec:defects}Defect motion}

\begin{figure}[t]
\centering
\includegraphics[width=\columnwidth]{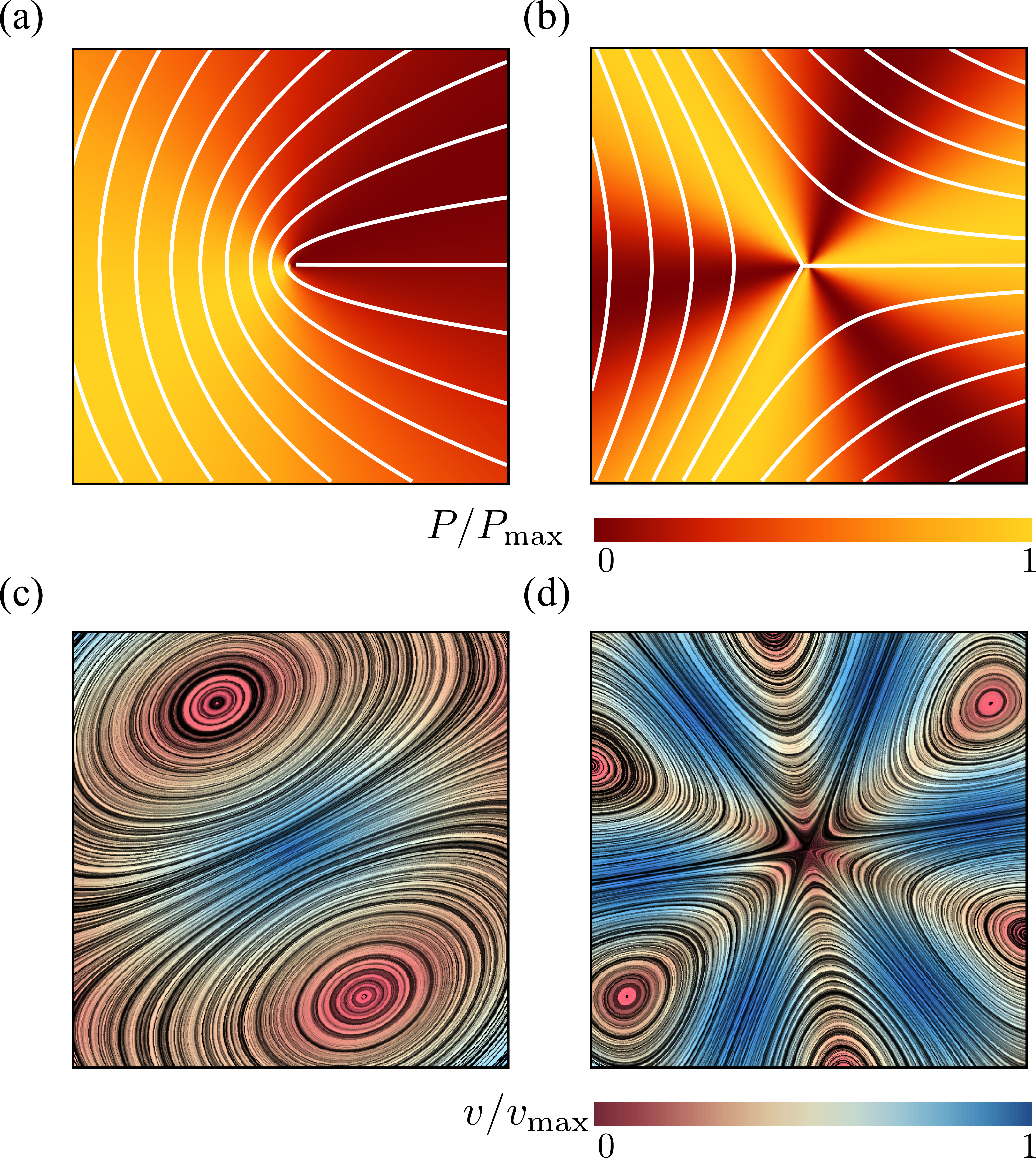}	
\caption{\label{fig:figure2}Pressure (a,b) and velocity (c,d) fields in proximity of $\pm 1/2$ defects obtained from the analytical solutions of Eq. \eqref{eq:stokes} for an extensile chiral active nematic with $\alpha=4\tau<0$. The configuration of the director is indicated by white lines for the case of $+1/2$ (a) and $-1/2$ (b) defects. For $+1/2$ defects, the velocity field at the core, thus the direction of motion of the defect, is tilted by an angle $\theta_{\rm tilt}=\arctan(1/2)\approx 27^{\circ}$ with respect to the defect polarity direction $\bm{p}=\bm{\hat{x}}$. For both chiral and achiral $+1/2$ defects, the pressure is anisotropic and larger toward the direction of motion of the defects.}
\end{figure}

The motion of $\pm 1/2$ disclinations has become a hallmark of active nematic liquid crystals. As it was theoretically predicted \cite{Giomi:2013,Giomi:2014} and experimentally verified in both microtubules-kinesin \cite{Sanchez:2012,Keber:2014,Guillamat:2016,Guillamat:2017} and actin-myosin suspensions \cite{Zhang:2018}, $\pm 1/2$ disclinations in active nematics drive a flow, whose structure and speed is strictly related to the geometry of the defect. For $+1/2$ disclinations, in particular, this flow has a Stokeslet structure that results in a propulsion of the defect in the direction of its symmetry axes $\bm{p}$ or $-\bm{p}$ depending on whether the system is contractile or extensile (Fig. \ref{fig:figure2} and Ref. \cite{Vromans:2016}).  Exceptionally, the same mechanism has been experimentally identified in various types of cell cultures, including spindle-shaped NIH 3T3 mouse embryo fibroblasts \cite{Duclos:2016}, murine neural progenitor cells (NPCs) \cite{Kawaguchi:2017}, Madin Darby canine kidney cells (MDCKs) \cite{Saw:2017} and human bronchial epithelial cell (HBEC) \cite{Blanch-Mercader:2018}. Unlike in suspensions of rod-like cytoskeletal filaments, however, the direction of motion of $+1/2$ disclinations is not necessarily parallel to $\pm\bm{p}$ and also the reconstructed flow around the defects is less symmetric than that found in cytoskeletal suspensions. Whereas it is not unlikely for such a feature to originate from statistical errors, here we demonstrate that chiral active stresses affect the dynamics of $+1/2$ defects precisely by rotating their direction of motion with respect to $\bm{p}$. In fact, the angle between $\bm{p}$ and the direction of motion of the defects could be used to indirectly measure the relative magnitude of the chiral stress $\tau$ compared to the deviatoric stress $\alpha$. This phenomenon shares some similarities with recent results by Maitra and Lenz about the dynamics of $+1/2$ disclinations in rotating active nematics \cite{Maitra:2019}.  

An analytical approximation of the flow driven by the active stresses in the surrounding of a disclination can be obtained by solving the incompressible Stokes equation with a body force resulting from the active stress associated with an isolated $\pm 1/2$ defect. Namely:
\begin{equation}\label{eq:stokes}
\eta\nabla^{2}\bm{v}+\bm{f}_{\pm}^{\rm a} = \nabla P\;,\qquad \nabla\cdot\bm{v} = 0\;,
\end{equation}
where $\eta$ is the shear viscosity of the tissue, here assumed isotropic for simplicity, and $\bm{f}_{\pm}^{\rm a}=\nabla\cdot\bm{\sigma}^{\rm a}$ is the body force arising from spatial variations of the active stress of Eq. \eqref{eq:active_stress} in the presence of a $\pm 1/2$ defect. This approach, introduced in Ref. \cite{Giomi:2014}, does not account for feedback of the flow on the orientation of the director and hence the structure of the defect. Nevertheless it provides a simple and faithful approximation of the defective flow as well as an estimate of the defect velocity. A generic solution of Eq. \eqref{eq:stokes} can be expressed in the form $\bm{v} = \bm{v}_{\pm}+\bm{v}_{0}$, where $\bm{v}_{0}$ is a solution of the homogeneous Stokes equation and is dictated by the boundary conditions and $\bm{v}_{\pm}$ is given by:
\begin{equation}\label{eq:v_pm}
\bm{v}_{\pm}(\bm{r}) = \int {\rm d}A'\,\bm{G}(\bm{r}-\bm{r}')\cdot\bm{f}_{\pm}^{\rm a}(\bm{r}')\;,
\end{equation}
where $\bm{G}$ is the two-dimensional Oseen tensor given by:
\begin{equation}\label{eq:oseen}
\bm{G}(\bm{r}) = \frac{1}{4\pi\eta} \left[\left(\log\frac{\mathcal{L}}{|\bm{r}|}-1\right)\bm{I}+\frac{\bm{r}\bm{r}}{|\bm{r}|^{2}}\right]\;,
\end{equation}
with $\mathcal{L}$ a constant set by the boundary conditions. Analogously, taking the divergence of Eq. \eqref{eq:stokes} allows to express the pressure field as the solution of the following Poisson equation:
\begin{equation}\label{eq:poisson}
\nabla^{2}P_{\pm} - \nabla\cdot\bm{f}_{\pm}^{\rm a} = 0\;.
\end{equation}%
Now, in the presence of a disclination of strength $s=\pm 1/2$ located at the origin of the $(x,y)-$plane and oriented in the direction $\bm{p}=(\cos\psi,\sin\psi)$, the nematic director $\bm{n}=(\cos\theta,\sin\theta)$ has local orientation $\theta=s\phi+(1-s)\psi$, with $\phi=\arctan (y/x)$ the polar angle \cite{Vromans:2016}. The body force $\bm{f}_{\pm}^{\rm a}$ is then readily found in the form:
\begin{widetext}
\begin{equation}\label{eq:defect_force}
\bm{f}_{\pm}^{\rm a}
= \frac{1}{2r}
\left\{
\begin{array}{lll}
\alpha\bm{p}+2\tau\bm{p}^{\perp}\;, & & \text{for } s=+1/2\;, \\[10pt]
-(\alpha \cos2\phi + 2\tau \sin 2\phi) \bm{p} + (\alpha \sin 2\phi - 2\tau \cos 2\phi) \bm{p}^{\perp}\;, & & \text{for } s = -1/2\;,
\end{array}
\right.
\end{equation}
\end{widetext}
where $\bm{p}^{\perp}=(-p_{y},p_{x})$ and $r$ is the distance form the defect core. The effect of chirality is most dramatic for $+1/2$ disclinations. In achiral active nematics, $\tau=0$ and the active force $\bm{f}_{+}^{\rm a}$ is purely longitudinal. For non-zero $\tau$, $\bm{f}_{+}^{\rm a}$ acquires a transverse component resulting in a tilt in the direction of motion of the defect by an angle
\begin{equation}\label{eq:tilt_angle}
\theta_{\rm tilt} = \arctan\left(\frac{2\tau}{\alpha}\right)
\end{equation}
with respect to the orientation $\bm{p}$ (Fig. \ref{fig:figure2}). As anticipated, Eq. \eqref{eq:tilt_angle} can in principle be used in combination with experimental reconstruction of defect trajectories in order to estimate the relative magnitude of the chiral and deviatoric stresses in nematic cell monolayers.

To calculate the flow velocity in proximity of a $\pm 1/2$ defect, we set, without loss of generality, $\psi=0$, thus $\bm{p}=\bm{\hat{x}}$ and $\bm{p}^{\perp}=\bm{\hat{y}}$ and we assume the defect at the center of a circular domain of radius $R$. The exact velocity of the flow at the boundary of such a domain, hence the homogeneous solution $\bm{v}_{0}$, is not relevant for the purpose of this discussion. In practice, this will be determined by the chemistry of the substrate and the possible presence of other topological defects in the same region \cite{Duclos:2016}. Under these assumptions, and carrying out algebraic manipulations as those in Ref. \cite{Giomi:2014}, the flow velocity caused by a $\pm 1/2$ defect can be found from Eqs. \eqref{eq:v_pm}, \eqref{eq:oseen} and \eqref{eq:defect_force} in the form:
\begin{subequations}\label{eq:defect_flow}
\begin{align}
\bm{v}_{+}
&= \frac{\alpha}{12\eta}\left\{[3(R-r)+r\cos 2\phi]\bm{\hat{x}}+ r\sin 2\phi\,\bm{\hat{y}}\right\} \notag\\[5pt]
&+ \frac{\tau}{6\eta}\left\{r\sin 2\phi\,\bm{\hat{x}} + [3(R-r)-r \cos 2\phi]\bm{\hat{y}}\right\}\;,\\[15pt]
\bm{v}_{-}
&= \frac{\alpha r}{12\eta R}
\Big\{
\left[\left(\frac{3}{4}r-R\right)\cos 2\phi-\frac{R}{5}\cos 4\phi\right]\bm{\hat{x}} \notag\\[5pt]
&-\left[\left(\frac{3}{4}r-R\right)\sin 2\phi+\frac{R}{5}\sin 4\phi\right]\bm{\hat{y}}\Big\}\notag\\[5pt]
&+\frac{\tau r}{6\eta R}
\Big\{
\left[\left(\frac{3}{4}r-R\right)\sin 2\phi-\frac{R}{5}\sin 4\phi\right]\bm{\hat{x}} \notag\\[5pt]
&+\left[\left(\frac{3}{4}r-R\right)\cos 2\phi+\frac{R}{5}\cos 4\phi\right]\bm{\hat{y}}\Big\} \;.
\end{align}
\end{subequations}
These flows are illustrated in Fig. \ref{fig:figure2}c,d for a specific choice of the angle $\theta_{\rm tilt}$. The corresponding pressure field is readily found from Eq.  \eqref{eq:poisson} with $\nabla\cdot\bm{f}_{+}^{\rm a}=-(\alpha \cos\phi+2\tau \sin\phi)/(2r^2)$ and $\nabla\cdot\bm{f}_{-}^{\rm a}=3(\alpha \cos 3\phi+2\tau \sin 3\phi)/(2r^2)$. This yields:
\begin{subequations}\label{eq:pressure}
\begin{align}
P_+ &= P_{+}^{(0)}+\frac{1}{2}\,(\alpha \cos \phi + 2\tau \sin \phi)\;,\\
P_- &= P_{-}^{(0)}-\frac{1}{6}\,(\alpha \cos 3\phi + 2\tau\sin3\phi)\;.
\end{align}
\end{subequations}
where $P_{\pm}^{(0)}$ are harmonic functions depending on the boundary conditions. Interestingly, the pressure field given by Eqs. \eqref{eq:pressure} is independent on the distance from the defect core, but varies with the angle $\phi$ and, depending on the sign of the stresses $\alpha$ and $\tau$, is maximal or minimal at specific directions relatively to the polarity vector $\bm{p}$. For instance, for extensile systems (i.e. $\alpha<0$) with negligible chirality (i.e. $\tau\approx 0$), Eq. (\ref{eq:pressure}a) predicts a pressure maximum in the $-\bm{p}$ direction, thus toward the ``head'' of the comet-like structure characteristic of $+1/2$ defects. Despite our analysis revolving around incompressible systems, we can expect this trend to persist in the presence of small spatial variations of the density field $\rho$. In this case, and under the assumption that $\rho \sim P$, we expect a higher density of extensile (contractile) cells in the front (back) of $+1/2$ defects, consistent with the experimental observation of Kawaguchi {\em et al.} \cite{Kawaguchi:2017}. Finally, the lack of spatial dependence in Eqs. \eqref{eq:pressure} originates for the specific parametrization of the orientation field $\theta$ in proximity of the defects, i.e. $\theta(\phi)=\pm \phi/2$. In a more realistic setting, $\theta(r,\phi)=\pm \phi/2+\theta_{\rm far}(r,\phi)$, where $\theta_{\rm far}$ is a non-singular function that determines the far-field configuration of the nematic director and vanishes at the defect core. Accounting for the far-field configuration of the director results into a dependence of the pressure on the distance from the defect core.

In summary, the presence of a symmetric chiral active stress, such as that embodied by the parameter $\tau$, affects the flow generated by $\pm 1/2$ disclinations by stretching and rotating the velocity field in the surrounding of the defects (Fig. \ref{fig:figure2}c,d). Most prominently, this results in a tilt in the direction of motion of $+1/2$ defects: i.e. $\bm{v}_{\rm self}=\bm{v}_{+}(r=0)=R/(4\eta)(\alpha\bm{p}+2\tau\bm{p}^{\perp})$. Thus $+1/2$ defects self-propel at an angle $\theta_{\rm tilt}$ with respect to their orientation $\bm{p}$ [see Eq. \eqref{eq:tilt_angle}]. Such an angle could in principle be measured in experiments on two-dimensional cell cultures, thus providing a direct measurement of the relative magnitude of the chiral stress. The same behavior has been reported in the case of actively rotating $+1/2$ defects, in the limit of vanishing angular velocity \cite{Maitra:2019}.

\section{\label{sec:adhesive_stripes}Spontaneous flow on adhesive stripes}

\begin{figure}[t]
\centering
\includegraphics[width=1\columnwidth]{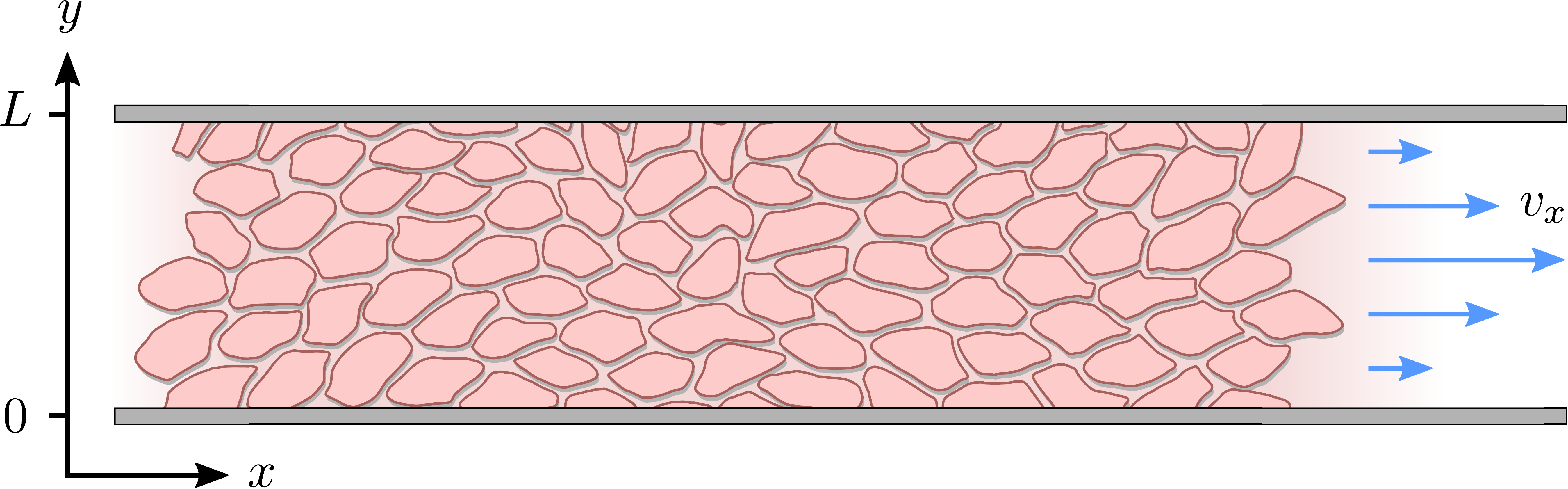}	
\caption{\label{fig:SketchChannel}Schematic representation of the channel of infinite length in $x$-direction and width $L$ in $y$-direction.}
\end{figure}

In this section we revisit a classic problem of the hydrodynamic stability and spontaneous flows of active nematics in a quasi-one-dimensional channel (Fig. \ref{fig:SketchChannel}). First discussed in a seminal paper by Voituriez {\em et al.} \cite{Voituriez:2005}, later elaborated by many others \cite{Marenduzzo:2007,Giomi:2008,Edwards:2009} and recently observed in experiments on cell monolayers \cite{Duclos:2018} and suspensions of microtubules and kinesin \cite{Hardouin:2019}, this phenomenon consists of a continuous transition between a stationary and uniformly oriented configuration to a state characterized by a spontaneous distortion of the nematic director coupled to an internally driven shear flow. The transition, in many aspects similar to the Fr\'eedericksz transition in passive liquid crystals \cite{Freedericksz:1927}, results as a consequence of two different mechanisms. First, a distortion of the nematic director drives a shear flow as illustrated by Eq. \eqref{eq:stokes} for the time-independent case; second, the nematic director rotates in a shear flow. As a consequence, when the hydrodynamic torque driven by the active stresses outweighs the elastic restoring torque, the uniformly oriented configuration becomes unstable to splay or bending deformations, depending on the sign of the active stress $\alpha$ and other material parameters. Roughly speaking, this occurs when the active length scale $\ell_{\rm a}=\sqrt{K/|\alpha|}$, defined by the ratio between the passive and active torques, with $K$ the Frank elastic constant, becomes comparable to the width $L$ of the channel \cite{Giomi:2015}. 

In the following, we extend and generalize the theoretical analysis by Duclos {\em et al}. \cite{Duclos:2018} by considering various experimentally relevant scenario in terms of boundary anchoring and flow. The hydrodynamic equations governing the dynamics of an incompressible (i.e. $\nabla\cdot\bm{v}=0$) active nematic liquid crystal are given by \cite{Voituriez:2005,Giomi:2008,Edwards:2009}:
\begin{subequations}\label{eq:constitutive1}
\begin{gather}
\frac{Dn_{i}}{Dt} = (\delta_{ij}-n_{i}n_{j})\left(\lambda u_{jk}n_{k}-\omega_{jk}n_{k}+\frac{1}{\gamma}\,h_{j}\right)\;, \\
\frac{Dv_{i}}{Dt} = \partial_{j}(-P\delta_{ij}+2\eta u_{ij}+\sigma_{ij}^{\rm e}+\sigma_{ij}^{\rm a})\;,
\end{gather}
\end{subequations}
where $D/Dt=\partial_{t}+\bm{v}\cdot\nabla$ is the material derivative, $u_{ij}=(\partial_{i}v_{j}+\partial_{j}v_{i})/2$ and $\omega_{ij}=(\partial_{i}v_{j}-\partial_{j}v_{i})/2$ are respectively the strain-rate and vorticity tensor and $\bm{h}=-\delta F/\delta \bm{n}$ is the so-called molecular field, governing the relaxational dynamics of the nematic director, with $F$ the Frank free energy. In one-elastic-constant approximation, the latter is simply given by $F=K/2\int {\rm d}A\,|\nabla\bm{n}|^{2}$ and $\bm{h}=K\nabla^{2}\bm{n}$. The material parameters $\lambda$ and $\gamma$ are respectively the flow-alignment parameter, controlling the tendency of the nematic director to rotate in a shear flow, and the rotational viscosity of the nematic fluid. In Eq. (\ref{eq:constitutive1}b), the pressure $P$ includes the active pressure $P^{\rm a}$ given in Eqs. \eqref{eq:active_stress} and \eqref{eq:microscopic}, but, due to incompressibility, is not an independent field and $\bm{\sigma}^{\rm e}$ is the elastic stress resulting from the departure of the director configuration from the ground state of the Frank free energy. Although it does affect the onset of the spontaneous flow instability, the latter is often unimportant for the phenomenology of active nematics and will be neglected here for simplicity.

We solve Eqs. \eqref{eq:constitutive1} in a rectangular channel which is infinitely long in $x-$direction and has width $L$ in $y-$direction (Fig. \ref{fig:SketchChannel}). 
Because the channel is invariant for translations along the longitudinal direction, we assume for simplicity $\bm{n}$ and $\bm{v}$ to be independent of $x$, both while stationary and spontaneously flowing. Furthermore, incompressibility and mass conservation demand the $y-$component of the velocity to vanish identically, i.e. $v_y = 0$. Thus taking $\bm{n} = (\cos \theta, \sin \theta)$ and $\bm{v}=({v_{x},0})$, the Eqs. \eqref{eq:constitutive1} reduce to:
\begin{subequations}\label{eq:dynamics}
\begin{gather}
\partial_t \theta= \frac{K}{\gamma}\partial_y^2 \theta - \frac{\partial_y v_x}{2} (1-\lambda \cos 2\theta)\;,\\
\partial_t v_x = \partial_y \left( \eta\partial_{y}v_{x} + \frac{\alpha}{2}  \sin 2\theta + \tau \cos 2\theta \right)\;,
\end{gather}
\end{subequations}
whereas from the $y-$component of Eq. (\ref{eq:constitutive1}b) we obtain the following expression for the pressure field:
\begin{equation}
P=P_0 - \frac{\alpha}{2} \cos 2 \theta + \tau \sin 2\theta\;,
\end{equation}
with $P_{0}$ a constant. Next, we look for stationary solutions of Eqs. \eqref{eq:dynamics} subject to different boundary conditions in terms of orientation of the nematic director at the boundary and whether or not the cells are allowed to slide along the channel walls, thus: $\theta=\theta(y)$, $v_{x}=v_{x}(y)$. In order for the fluid to be stationary, the shear stress must be uniform across the channel. Thus, from Eq. (\ref{eq:dynamics}b):
\begin{equation}\label{eq:sigma_constant}
\sigma_{xy} = \eta \partial_{y}v_{x}+\frac{\alpha}{2}\sin 2\theta + \tau \cos 2\theta = {\rm const}\;.	
\end{equation}
Solving Eq. \eqref{eq:sigma_constant} with respect to $\partial_{y}v_{x}$ and substituting this into Eq. (\ref{eq:dynamics}b) yields a single homogeneous equation for $\theta$, namely:
\begin{equation}\label{eq:theta}
\frac{K}{\gamma}\partial_{y}^{2}\theta=\frac{1}{2\eta}\left(\sigma_{xy}-\frac{\alpha}{2}\sin 2\theta-\tau\cos 2\theta \right)(1-\lambda\cos 2\theta) \;. 
\end{equation}
Before analyzing specific cases in detail, we consider the generic situation in which the nematic director is anchored to the channel walls at an arbitrary angle $\theta_{0} \ne \arccos(1/\lambda)/2$, thus:
\begin{equation}\label{eq:theta_bc}
\theta (0) = \theta (L) = \theta_{0}\;.
\end{equation}
In the following, we will separately analyze the scenarios in which the cells are stationary at the boundary of the channel (Sec. \ref{sec:4A}) and when, on the other hand, they are able to slide while keeping their orientation fixed (Sec. \ref{sec:4B}).

Our analysis is complemented by numerical solutions of Eqs. \eqref{eq:dynamics} with various boundary conditions. For this purpose we rescale time by the viscous time scale $\tau_\nu =\rho L^2/\eta$, length by the channel width $L$ and stress by the viscous stress scale $\sigma_\nu = \rho L^2/\tau_\nu^2$, i.e. $t \to t/\tau_\nu$, $y \to y/L$, $\sigma \to \sigma/\sigma_\nu$. All the other quantities in Figs. \ref{fig:BifE} and \ref{fig:PlotsFig5} are rescaled accordingly.

\subsection{\label{sec:4A}No-slip boundary conditions}

In this section, we consider the case in which the cells are unable to slide along the boundary of the channel, i.e. $v_{x}(0)=v_{x}(L)=0$, which, in turn, experience a non-vanishing stress $\sigma_{xy}\ne0$ resulting from the cellular forces. In this case, $\theta(y)=\theta_{0}$ is always a trivial solution of Eq. \eqref{eq:theta} with the boundary condition given by Eq. \eqref{eq:theta_bc}, and the monolayer admits a stationary and uniformly aligned configuration. Because of the internal stresses, however, such a uniform state can become unstable with respect to splay or bending deformations for sufficiently large active stresses or channel widths. To illustrate this point, we take $\theta(y)=\theta_{0}+\delta\theta(y)$, with $\delta\theta(0)=\delta\theta(L)=0$, and linearize Eq. \eqref{eq:theta} about $\delta\theta=0$. This yields:
\begin{equation}
\label{eq:LinThetaA}
\partial_y^2 \delta \theta + q^2 \delta \theta = 0 \;,
\end{equation}
where we have introduced the constant:
\begin{align}\label{eq:q2}
q^2  
&= \frac{\gamma}{\eta K}\Big[\left(1-\lambda \cos 2\theta_{0}\right)\left(\frac{\alpha}{2}\,\cos 2\theta_{0}-\tau \sin 2\theta_{0}\right) \notag \\
&- \lambda \sin 2\theta_{0}\left(\sigma_{xy}-\frac{\alpha}{2}\,\sin 2\theta_{0}-\tau \cos 2\theta_{0}\right)\Big]\;.
\end{align}
The solution of Eq. \eqref{eq:LinThetaA} is readily found to be:
\begin{equation}
\label{eq:SolLinThetaA1}
\delta \theta = C \sin(q y) \;, \quad q L = n \pi\;,
\end{equation}
with a constant $C$ and $n\in\mathbb{Z}$ an arbitrary integer. By virtue of Eq. \eqref{eq:sigma_constant}, the corresponding velocity field is given by:
\begin{equation}
\label{eq:VelocLinA1}
\eta \partial_y v_x = \delta \theta \left(2 \tau \sin 2\theta_{0} - \alpha \cos 2\theta_{0} \right) \;.
\end{equation}
whose solution with no-slip boundary conditions is given by:
\begin{equation}\label{eq:SolLinVelA1}
v_x = \frac{C}{q\eta}\,(2\tau \sin 2\theta_{0} - \alpha\cos 2\theta_{0})(1-\cos qy) \;, \quad q L = 2 m \pi\;, 
\end{equation}
with $m \in \mathbb{Z}$ another arbitrary integer. Upon comparing Eqs. \eqref{eq:SolLinThetaA1} and \eqref{eq:SolLinVelA1} we find that the first mode to be excited is $(n,m)=(2,1)$, thus the trivial solution $\theta(y)=\theta_{0}$ becomes unstable when $q=q_{c}=2\pi/L$ or, equivalently, when $L=L_{c}=2\pi/q$. To be more specific, we consider, in the following, two practically relevant cases, where $\theta_0=0$ (parallel anchoring) and $\theta_{0}=\pi/2$ (homeotropic anchoring).

\subsubsection{\label{sec:4A1}Parallel anchoring}

If the nematic director is parallel to the channel walls $\theta_{0}=0$, thus:
\begin{equation}\label{eq:q2_parallel}
q^2 = q_{\parallel}^{2} =  \frac{\alpha \gamma (1-\lambda)}{2 \eta K}\;.
\end{equation}
As in non-chiral active nematics, the instability is triggered by splay deformations (i.e. transverse to the nematic director) and uniquely depends on the non-chiral active stress $\alpha$. Furthermore, as in non-chiral active nematics \cite{Giomi:2014}, such a splay instability affects flow-aligning systems (i.e. $\lambda>1$) in the presence of extensile active stresses (i.e. $\alpha<0$), and flow-tumbling systems (i.e. $\lambda<1$) in the presence of contractile active stresses (i.e. $\alpha>0$). A critical $\alpha$ value is readily found in the form: $\alpha_{c}=8\pi^{2}\eta K /[\gamma L^{2}(1-\lambda)]$. 

At the onset of the transition, the constant $C$ can be calculated upon expanding Eq. \eqref{eq:theta} up to the third order in $\delta\theta$, Then, using the solution of the linearized equation yields a cubic equation in $C$. Solving the latter gives (see Appendix \ref{sec:appendixB}):
\begin{gather*}
\delta\theta \approx \pm \sqrt{\left(\frac{L}{L_c}-1\right)\left[\frac{3}{4+3/(\lambda-1)}\right]}\sin \frac{2\pi y}{L},\\[7pt]
v_x \approx \pm \frac{\alpha}{\eta q_\parallel} \sqrt{\left(\frac{L}{L_c}-1\right)\left[\frac{3}{4+3/(\lambda-1)}\right]}\left( \cos \frac{2\pi y}{L} - 1\right).
\end{gather*}
\begin{figure}[t]
\centering
\includegraphics[width=\columnwidth]{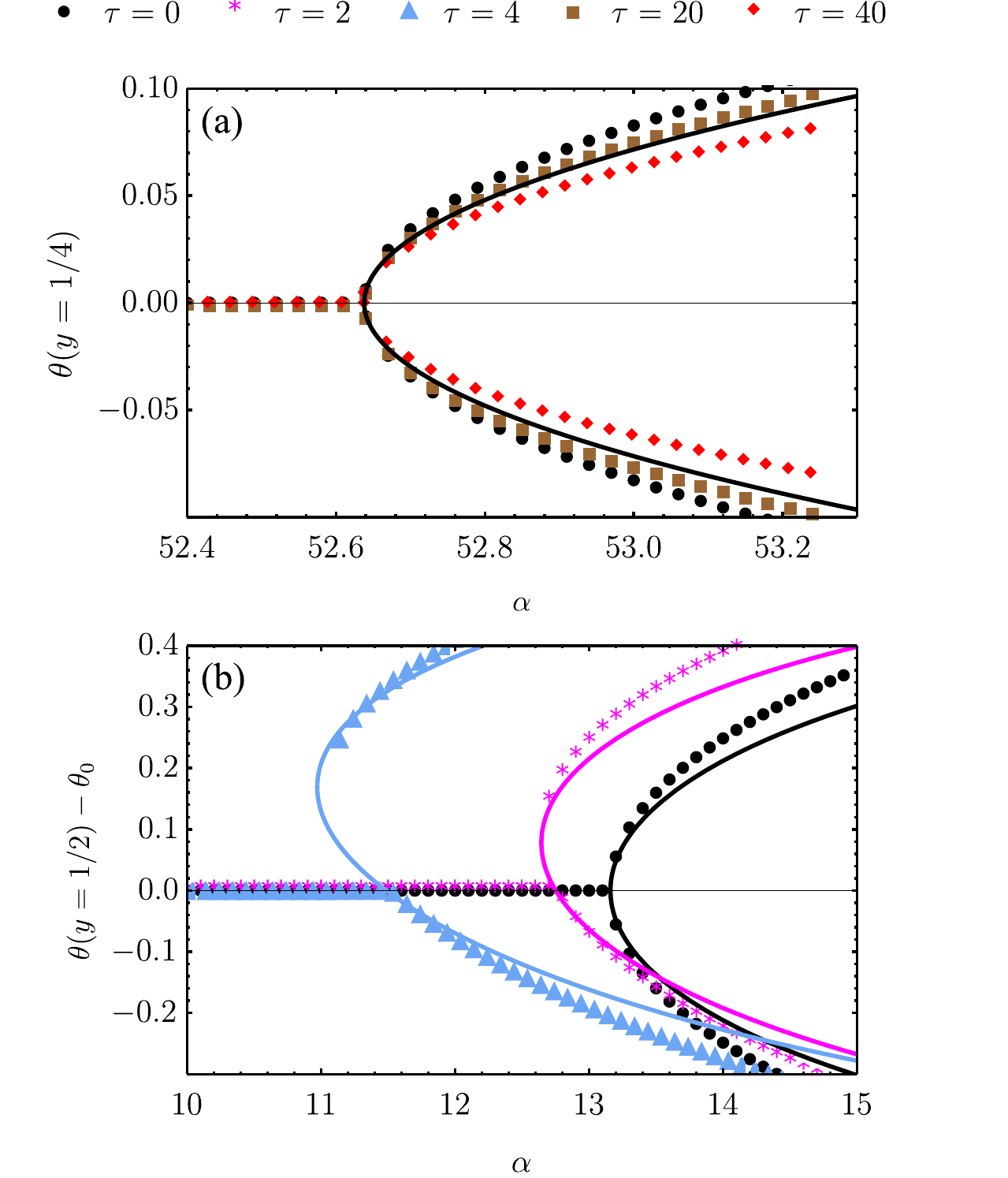}
\caption{\label{fig:BifE}Bifurcation diagram of the spontaneous flow transition obtained from numerical (dots) and analytical (lines) solutions of Eqs. \eqref{eq:dynamics} for $K/\gamma=L=1$, $\lambda = -0.5$ and various $\tau$ values. (a) No-slip boundary conditions and parallel anchoring (see Sec. \ref{sec:4A1}). The chiral stress $\tau$ does not influence the critical $\alpha$ value, but weakly affect the post-transitional configuration of the nematic director. (b) Stress-free boundary conditions and special boundary anchoring $\theta(0)=\theta(L)=-\theta_{\rm tilt}/2$ (see Sec. \ref{sec:4B3}). The chiral active stresses, embodied by the parameter $\tau$, explicitly break the clock-counterclockwise symmetry of the lowest free-energy configuration rendering the pitchfork bifurcation ``imperfect''. In this case, only one of the two branches of the bifurcation diagram is connected to the trivial solution, which may then be the only one observed experimentally.}
\end{figure}
The spontaneous flow instability consists, therefore, of a standard pitchfork bifurcation whose relevant fields, $\theta$ and $v_{x}$, scale like $(L-L_{c})^{1/2}$ at criticality, see Fig \ref{fig:BifE}a.  Despite the fact that the chiral stress $\tau$ does not affect the instability of the stationary state, it leaves a clear signature on the post-transitional behavior of the flowing monolayers. This can be seen in Fig. \ref{fig:PlotsFig5}a, showing numerical solutions of Eqs. \eqref{eq:dynamics} in the flowing state for various $\alpha$ and $\tau$ values. The most prominent effect of chirality, in this case, is evidently to render both the distortion of the nematic director and the associated flow asymmetric with respect to the channel centerline.

\begin{figure*}[htb]
    \centering
  \includegraphics[width=\textwidth]{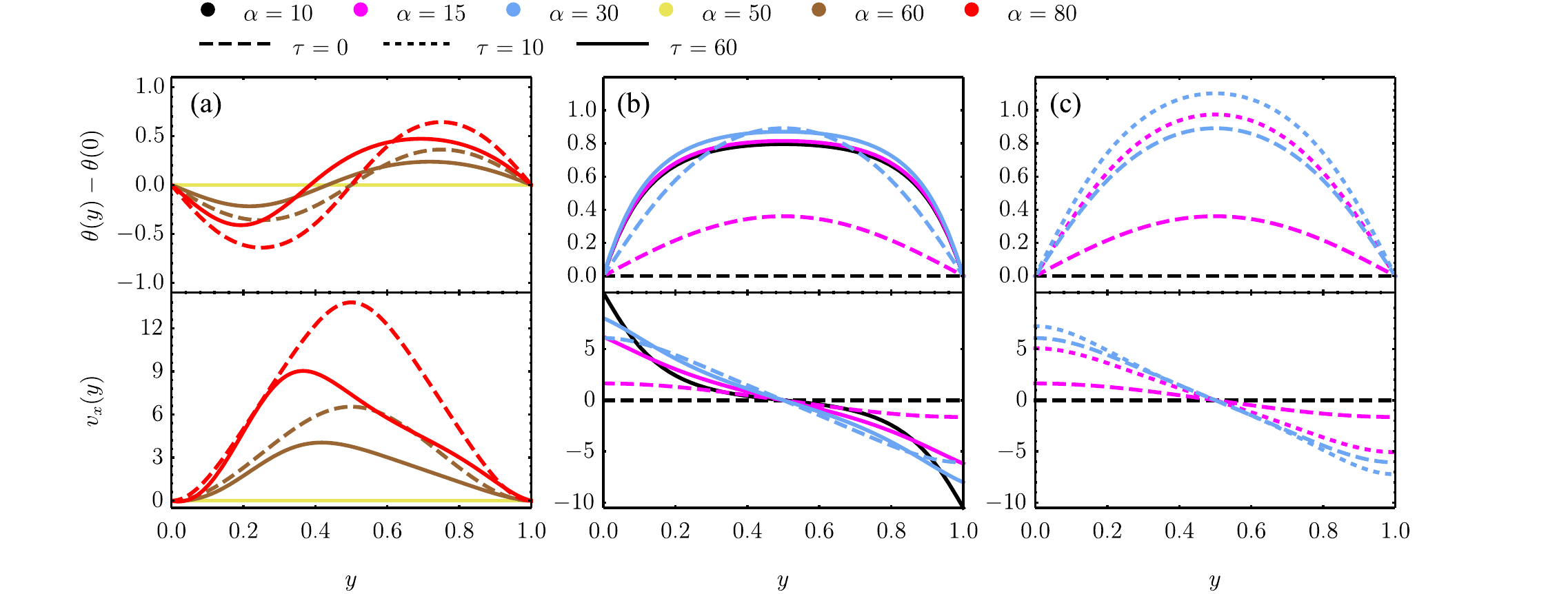}
  \caption{\label{fig:PlotsFig5}Numerical solutions of Eqs. \eqref{eq:dynamics} in the dimensionless quantities defined above and for $K/\gamma=1$, $\lambda = -0.5$ as well as different values of $\tau$ and $\alpha$ for three sets of boundary conditions. (a) and (b) are numerical solutions for the Eqs. \eqref{eq:dynamics} with parallel anchoring as well as no-slip boundary conditions [panel (a) and analyzed analytically in Sec. \ref{sec:4A1}] and stress-free boundary conditions [panel (b) and Sec. \ref{sec:4B1}], respectively. (c) Numerical solutions for stress-free walls and the stationary solution $\theta_0 = -\theta_\text{tilt}/2=-\arctan(2\tau/\alpha)/2$ imposed at the boundary, see Sec. \ref{sec:4B3}.}
\end{figure*}
\subsubsection{\label{sec:4A2}Homeotropic anchoring}

The case in which the nematic director is perpendicularly anchored to the channel walls, $\theta_{0}=\pi/2$, yields: 
\begin{equation}\label{eq:q2_homeotropic}
q^{2} = q_{\perp}^{2} = -\frac{\alpha\gamma(1+\lambda)}{2\eta K}\;. 	
\end{equation}
In this case the instability is triggered by bending deformations (i.e. parallel to the nematic director). In contrast with the scenario of Sec. \ref{sec:4A1}, flow-aligning systems are unstable in the presence of extensile active stresses, whereas strongly flow-tumbling systems (i.e. $\lambda<-1$), are unstable in the presence of contractile active stresses. The critical $\alpha$ value is readily found in the form: $\alpha_{c}=-8\pi^{2} \eta K/[\gamma L^{2}(1+\lambda)]$. 

To conclude this section, we observe that for both parallel and homeotropic anchoring, the spontaneous flow instability crucially relies on the flow-alignment behavior of the system, governed by the phenomenological parameter $\lambda$. As for molecular liquid crystals, where $\lambda$ depends upon the molecules shape and interactions, we expect the flow-alignment parameter to be affected by the cellular shape, which in turn is not fixed, and by the passive and active processes underlying the cell-cell and cell-substrate interactions.

\subsection{\label{sec:4B}Stress-free boundary conditions}

In this subsection we consider the scenario in which the cells are allowed to slide along the boundary, while keeping a fixed orientation $\theta_{0}$ with respect to the channel walls. As in the adhesive stripes used in Ref. \cite{Duclos:2018}, the channel walls do not comprise a real physical barrier, but represent instead the interface between two regions of the substrate with different coating. As the walls now do not exert any force on the cells $\sigma_{xy}(0)=\sigma_{xy}(L)=0$. Mechanical equilibrium [i.e. Eq. \eqref{eq:sigma_constant}] thus implies $\sigma_{xy}=0$ everywhere. 

The most striking difference with respect to the case discussed in Sec. \ref{sec:4A}, as well as the most prominent consequence of the chiral stress $\tau$, is that the stationary and uniformly aligned configuration (i.e. $\theta=\theta_{0}$ and $v_{x}=0$) is not a trivial solution of Eq. \eqref{eq:theta} for non-vanishing $\alpha$ and $\tau$ values, unless the chiral and non-chiral active stresses cancel each other identically. Before considering this latter case (see Sec. \ref{sec:4B3}), we find approximated expressions for the local orientation $\theta$ and the velocity $v_{x}$ in the limits in which the active stresses are either very small or very large.

For very large active stresses, the active terms in Eq. \eqref{eq:theta} overweight the elastic term on the left hand side. As a consequence, the equilibrium configuration of the nematic monolayer consists of a region in the bulk of the channel where the cells have uniform orientation $\theta=-\theta_{\rm tilt}/2=-\arctan(2\tau/\alpha)/2$ and two boundary layers, whose size is roughly $\xi_{\rm bl} \sim \sqrt{K/(\gamma \sigma_{0})}$, with $\sigma_{0}=(\alpha/2)\sin 2\theta_{0}+\tau\cos 2\theta_{0}$, where the director interpolates between the bulk and boundary orientation (see Fig. \ref{fig:PlotsFig5}b). This phenomenon closely resembles flow-alignment in nematics (see e.g. Ref. \cite{DeGennes:1993}) with $-\theta_{\rm tilt}/2$ playing the role of the Leslie angle $\theta_{L}=\arccos(1/\lambda)/2$. Whereas passive flow-alignment, however, requires $\lambda>1$ (e.g. flow-aligning nematics), such an {\em active flow-alignment} occurs at any finite value of $\alpha$ and $\tau$, provided the elastic boundary layer is sufficiently small to have a clear distinction between bulk and boundary alignment.

For small $\alpha$ and $\tau$ values, we can postulate that the nematic director will depart only slightly from its orientation at the boundary. Thus, taking again $\theta(y)=\theta_{0}+\delta\theta(y)$ and linearizing Eq. \eqref{eq:theta} around $\delta\theta=0$, we obtain:
\begin{equation}\label{eq:LinThetaB}
\partial_{y}^{2}\delta\theta + q^{2}(\delta\theta+\delta\theta_{0})=0\;,
\end{equation}
with $q^{2}$ given, as before, by Eq. \eqref{eq:q2} and:
\begin{equation}\label{eq:delta_theta_0}
q^{2}\delta\theta_{0} = \frac{\gamma}{2 \eta K}\,(1-\lambda \cos 2\theta_{0})\left(\frac{\alpha}{2}\,\sin 2\theta_{0}+\tau \cos 2\theta_{0}\right)\;.
\end{equation}
For a general anchoring angle $\theta_{0}$, a solution of Eq. \eqref{eq:LinThetaB} with boundary conditions $\delta\theta(0)=\delta\theta(L)=0$ is given by:
\begin{equation}
\label{eq:SolLinThetaA2}
\delta \theta = \delta \theta_{0} \left(\cos qy + \sin qy \tan \frac{q L}{2} -1 \right)\;.
\end{equation}
The associated velocity field can then be found from a direct integration of the linearized equation:
\begin{multline}
\label{eq:VelocLinA2}
\eta \partial_y v_x + \left(\alpha \cos 2\theta_{0} - 2 \tau \sin 2\theta_{0}\right)\delta \theta \\
+ \frac{\alpha}{2} \sin 2\theta_{0} + \tau \cos 2\theta_{0} = 0 \,.
\end{multline}
The lack of a boundary condition for Eq. \eqref{eq:VelocLinA2} can be compensated with a global constraint on the total momentum, namely $\int_{0}^{L}\dd y\, v_{x}=0$. 

In the following, we provide explicit approximated expression for the velocity field in the special cases where $\theta_{0}=0$ (parallel anchoring) and $\theta_{0}=\pi/2$. Furthermore, we will investigate the stability of the trivial solution of Eq. \eqref{eq:theta} obtained when $\theta_{0}$ is such that the chiral and non-chiral active stresses cancel each other identically.

\subsubsection{\label{sec:4B1}Parallel anchoring}
For $\theta_{0}=0$
Eqs. \eqref{eq:q2_parallel} and \eqref{eq:delta_theta_0} yield:
\begin{equation}\label{eq:delta_0}
\delta\theta_{0} = \frac{\tau}{\alpha}\;.
\end{equation}
The corresponding velocity field is then readily obtained by integrating Eq. \eqref{eq:VelocLinA2}. This gives:
\begin{equation}\label{eq:vx}
v_x \approx -\frac{\tau \sin q_{\parallel} y}{q_{\parallel} \eta} + \frac{\tau \cos q_{\parallel} y \tan\frac{q_{\parallel} L}{2}}{q_{\parallel} \eta}\;,
\end{equation}
where the wave number $q_{\parallel}$ is that given in Eq. \eqref{eq:q2_parallel}. Numerical solutions for this case are displayed in Fig. \ref{fig:PlotsFig5}b for various $\alpha$ and $\tau$ values.

We stress that, whereas the flowing configurations resulting from the instability of the stationary state are left-right and clock-counterclockwise symmetric (i.e. the cells are equally likely to flow toward the negative or positive $x-$direction and, correspondingly, to tilt clock- or counterclockwise, see Sec. \ref{sec:4A1}), in this case the direction of the tilt as well as that of the flowing monolayer is set by the signs of the constants $\alpha$ and $\tau$.

\subsubsection{\label{sec:4B2}Homeotropic anchoring}
For $\theta_{0}=\pi/2$
from Eqs. \eqref{eq:q2_parallel} and \eqref{eq:delta_theta_0} we find that the amplitude $\delta\theta_{0}$ is given, once again, by Eq. \eqref{eq:delta_0}. Thus, the expressions for $\delta\theta$ and $v_{x}$ are formally identical to those given in Eqs. \eqref{eq:SolLinThetaA2} and \eqref{eq:vx}, but with wave number $q_{\perp}$ as given in Eq. \eqref{eq:q2_homeotropic}. 

\subsubsection{\label{sec:4B3}Stationary Solution}

To conclude this subsection, we consider a special situation where the orientation of the cells at the boundary is fixed, as before, but such that the chiral and non-chiral stresses cancel each other identically. Thus: $\theta_{0}=-\theta_{\rm tilt}/2=-\arctan(2\tau/\alpha)/2$. In this case the orientation of the nematic director in the bulk of the channel, determined by the balance between the chiral and non-chiral active stress, is equal to that at the boundary. As a consequence, the boundary layer described in Sec. \ref{sec:4B} disappears and the system can achieve a stationary and uniformly aligned configuration. As those described in Sec. \ref{sec:4A}, however, the latter is unstable for sufficiently large active stresses or channel width. 

Using the same algebraic manipulations adopted in Sec. \ref{sec:4A}, one can show that the perturbation $\delta\theta$ is again of the form given in Eq. \eqref{eq:SolLinThetaA1} with:
\begin{equation}
q^{2} = \frac{\gamma \left(\sqrt{\alpha^2+4\tau^2} - \alpha \lambda\right)}{2 \eta K}\;.
\end{equation}
Analogously, the velocity is given by Eq. \eqref{eq:SolLinVelA1}, but with no constraint on the phase $qL$, because of the stress-free boundary conditions. As a consequence, the first mode to be excited is $n=1$, thus the stationary state becomes unstable when $q=q_{c}=\pi/L$ or, equivalently, when $L=L_{c}=\pi/q$. Some numerical solution of Eq. \eqref{eq:dynamics}, in this regime, are shown in Fig. \ref{fig:PlotsFig5}c. 

Notably, the transition from stationary to flowing is, in this case, no longer left-right and clock-counterclockwise symmetric, as in the examples discussed in Sec. \ref{sec:4A}, for any $\tau \ne 0$. This is well illustrated by the bifurcation diagram of Fig. \ref{fig:BifE}b, showing the departure in the director orientation from the boundary value at the center of the channel [i.e. $\theta(1/2)-\theta_{0}$, with $L=1$]. The dots have been obtained from a numerical integration of Eqs. \eqref{eq:dynamics}, whereas the solid lines correspond to analytical solutions obtained by solving a third order equation for the constant $C$ in Eq. \eqref{eq:SolLinThetaA1}, as in Sec. \ref{sec:4A1}. We find:
\begin{equation}
\label{eq:AnalyticStatSol}
\theta(L/2)-\theta_{0} = \frac{\sqrt{6}\,\sin \frac{q_cL}{2}\left(\sqrt{6}\,\lambda \tau \pm \sqrt{a + b\,\frac{L_c^2}{L^2}} \right)}{2(4\alpha\lambda - \sqrt{\alpha^2+4\tau^2})} \;,
\end{equation}
where $a = \alpha^2 (1+4\lambda^2) + 2\tau^2(2+3\lambda^2)-5\alpha\lambda\sqrt{\alpha^2+4\tau^2}$ and $b = \eta K q^{2}(8 \alpha \lambda - 2 \sqrt{\alpha^2+4\tau^2})/\gamma$. For $\alpha>\alpha_{c}$, the solution consists of two branches, of which only one is connected with the stationary solution $\theta(y)=\theta_{0}$. Furthermore, the gap between the two branches increases monotonically with $\tau$. If the instability is triggered upon applying a small random perturbation to the stationary state, this will always select the closest branch, thus the one connected to the trivial solution. As a consequence, a chiral cellular monolayer driven out of the stationary state by a small perturbation, will systematically tilt and flow in the same direction, which is in turn determined by the sign of the chiral active stress $\tau$. The transition described above is known in bifurcation theory as a perturbed or imperfect pitchfork bifurcation and occurs when a standard pitchfork bifurcation, whose normal form is $\theta^3-\mu \theta=0$, is biased by a small symmetry-breaking perturbation: i.e. $\theta^3-\mu \theta + P_L + P_S \theta^2=0$, where $\mu$, $P_L$, and $P_S$ are constant parameters. If $P_{L}=P_{S}=0$, the equation is invariant under $\theta \to -\theta$. Thus, for $\mu > 0$, the trivial solution is unstable and the transition is supercritical, while for $\mu < 0$, only the trivial solution is stable, and the bifurcation is subcritical. By contrast, for non-vanishing $P_{L}$ and $P_{S}$, the equation is no longer invariant under $\theta \to -\theta$ and the bifurcation is no longer symmetric (see Refs. \cite{Golubitsky:1985,Kitanov:2013} for an overview).

In our case, the role of the symmetry-breaking perturbation is played by the chiral stress $\tau$. Thus, in the unperturbed scenario, $\tau = 0$ and the stationary solution is $\theta = 0$, with the critical $\alpha$ value being $\alpha_c = 2\pi^2\eta K/[\gamma L^2 (1-\lambda)]$. When $\tau \ne 0$, on the other hand, expanding Eq. \eqref{eq:theta} around $\theta=0$ and using $\partial_y^2\theta = -q^2\theta$ one finds:
\begin{multline}
\frac{2}{3}\,\alpha_c(4\lambda-1)\theta^3 - \left[\alpha(\lambda-\tau-1)+\sqrt{\alpha^2+4\tau^2}\right]\theta \\  
+\tau(1-\lambda) + 2\tau(2\lambda-1)\theta^2 + \mathcal{O}\left[(\alpha-\alpha_c)^2\theta^3\right] = 0 \;.
\end{multline}
Evidently, this coincides with the normal form of a perturbed bifurcation for any finite $\tau$ value. For $\tau=0$, on the other hand, one recovers the normal form of the symmetric pitchfork bifurcation. Upon increasing $\tau$, the bifurcation is shifted toward smaller $\alpha$ values, until, for $\tau = \pi^2/ \eta K  (\gamma L^2)$, $\alpha_{c}=0$ and the system is never stationary.

\section{\label{sec:conclusion}Discussion and conclusions}

In this article we have investigated how a chiral and yet symmetric stress tensor might arise microscopically in nematic cell monolayers and how such a chiral stress influences some of the hallmark phenomena of active nematics. In Sec. \ref{sec:chiral_stresses} we proposed a microscopic model where a misalignment of the active force dipole and the cell's orientation is demonstrated to lead to a macroscopic chiral active stress tensor of the form Eq. \eqref{eq:active_stress}. In Sec. \ref{sec:defects} we showed how the presence of chiral active stresses tilts the flow around $\pm 1/2$ disclinations, thereby leading to a misalignment between the defect polarity and the direction of motion, by an angle $\theta_{\rm tilt}=\arctan(2\tau/\alpha)$, with $\tau$ and $\alpha$ the chiral and non-chiral active stress respectively. In Sec. \ref{sec:adhesive_stripes} we investigated the spontaneous flow transition in a quasi-one-dimensional channel for both no-slip and stress-free boundary conditions as well as for various types of anchoring. For no-slip boundaries (Sec. \ref{sec:4A}), we recovered the classic pitchfork bifurcation first discussed by Voituriez {\em et al.} \cite{Voituriez:2005}. In this case the chirality does not affect the transition itself, but does leave a signature on the post-transitional configurations of the nematic director and velocity field, in the form of asymmetry with respect to the channel centerline. In case of stress-free boundaries (Sec. \ref{sec:4B}), we found that chirality renders the stationary and uniformly aligned configuration incompatible with most of the anchoring conditions. As a consequence, the cellular monolayer is always in motion, for any non-vanishing chiral and non-chiral active stress. For very large active stresses, in particular, we found an active analog of flow-alignment in nematics, with the bulk orientation (analogous to the Leslie angle \cite{DeGennes:1993}) set by the ratio between chiral and non-chiral active stresses, i.e. $-\theta_{\rm tilt}/2$. Finally, in the special case in which the nematic director is anchored at an angle $-\theta_{\rm tilt}/2$ at the channel walls, we found that the spontaneous flow transition becomes asymmetric, i.e. only one of the two branches of the pitchfork bifurcation is connected to the trivial solution, which may then be the only one observed experimentally. This latter result could potentially explain the experimental observations by Duclos {\em et al}. \cite{Duclos:2018}, who found that NIH 3T3 cells are more likely to tilt clockwise then counterclockwise once the spontaneous flow transition sets up.
In Ref. \cite{Blow:2014, Doostmohammadi:2016} it was found that at the interface between an active nematic phase, with negligible distortions of the director field, and an isotropic phase there is an active anchoring angle that is set by the activity. This discussion can easily be repeated for the chiral active stress tensor in Eq. \eqref{eq:active_stress} with the result being that the angle is changed due to the presence of chirality. Furthermore, considering a confined nematic phase without isotropic phase but with distorted director field, the case considered in Sec. \ref{sec:adhesive_stripes} being one example, it is readily found that the active anchoring angle is equal to the angle $\theta_{\rm tilt}$ introduced above. Thus, the active anchoring angle considered in Ref. \cite{Blow:2014, Doostmohammadi:2016} (no distortions of director field but region with active nematic and isotropic phase) and the anchoring angle $\theta_{\rm tilt}$ derived above (distorted director field but only a nematic phase) can be seen as being the two special cases of the more general case with a nematic-isotropic phase interface and non-negligible distortions of the director field.

Further experimental investigations into the influence of chirality would be interesting. In particular, the tilt of the flow around $\pm1/2$ disclinations has, according to our knowledge, not yet been observed. Thus, measurements of the tilt angle and experimental investigations of the flow field are needed to compare the theory with real-life cell monolayers. Additionally, as mentioned, the tilt angle opens a possibility to determine the relative magnitude of the chiral stress directly by particle-image-velocimetry measurements.
Furthermore, since the cells used in Ref. \cite{Duclos:2018} were only weakly chiral the effects of chirality were not as pronounced. Performing similar experiments with cells with stronger chirality and for different boundary conditions would enable further tests of the presented theory. 

\acknowledgments

We are grateful to Mattia Serra for illuminating discussions about Sec. IV. This work was supported by funds from the Netherlands Organisation for Scientific Research (NWO/OCW), as part of the Frontiers of Nanoscience program (L.G.), the Vidi (L.G. and L.A.H.), and Vici (R.M.H.M.) schemes and the Leiden/Huygens fellowship (K.S.).

\appendix 

\section{\label{sec:appendix}Stress owing to a force quadrupole}

The derivation of the active stresses given in Sec. \ref{sec:chiral_stresses} can be straightforwardly generalized to account for a more complex force distribution. For illustrative purposes we consider here the case of a quadrupole consisting of two force dipoles applied at the ends of the cell in longitudinal and transverse directions. In this case the force density field $\bm{f}_{c}$ is given by:
\begin{align}
\bm{f}_{c} 
&= \bm{F}_{c}^{(a)}\delta(\bm{R}_{c}-a\bm{\nu}_{c})-\bm{F}_{c}^{(a)}\delta(\bm{R}_{c}+a\bm{\nu}_{c})\nonumber \\	
&+ \bm{F}_{c}^{(b)}\delta(\bm{R}_{c}-b\bm{\nu}_{c}^{\perp})-\bm{F}_{c}^{(b)}\delta(\bm{R}_{c}+b\bm{\nu}_{c}^{\perp})\;,	
\end{align}
where $a$ and $b$ are, respectively, the major and minor semiaxis. Taking $\bm{F}_{c}^{(i)}=F_{\parallel}^{(i)}\bm{n}+F_{\perp}^{(i)}\bm{n}^{\perp}$, with $i=a,b$, and coarse-graining over the scale of a volume element, we find: $P^{\rm a}=-(aF_{\parallel}^{(a)}+bF_{\perp}^{(b)})$, $\alpha = 2(aF_{\parallel}^{(a)}-bF_{\perp}^{(b)})$ and $\tau = aF_{\perp}^{(a)}+bF_{\parallel}^{(b)}$.

\section{\label{sec:appendixB}Nonlinear Expansion}
Expanding Eq. \eqref{eq:theta} about $\theta_0 = 0$ up to third order in $\delta\theta$ yields
\begin{align*}
-\frac{K \eta}{\gamma}\partial_y^2 \delta\theta &= \frac{\alpha (\lambda-1) \delta\theta}{2} - \tau (\lambda-1) \delta\theta^2 + \frac{\alpha (1-4\lambda) \delta\theta^3}{3} \nonumber \\
&+\frac{(\tau-\sigma_{xy})(\lambda-1)}{2}\;,
\end{align*}
which can be written as
\begin{equation}
\frac{L_c^2}{L^2} \delta \theta = \delta \theta - \frac{2 \tau \delta\theta^2}{\alpha} + \frac{2 (1-4\lambda) \delta\theta^3}{3 (\lambda - 1)} +\frac{(\tau-\sigma_{xy})}{\alpha} \;.
\end{equation}

The stress $\sigma_{xy}$ is determined by the no-slip boundary condition and force balance, i.e.,
\begin{equation}
\label{eq:ThirdOrderDeltaEq}
\int_0^L dy^\prime \left(\sigma_{xy}-\frac{\alpha}{2}\sin2\theta-\tau\cos2\theta\right) = 0 \;.
\end{equation}
An expansion up to third order in $\delta\theta = C \sin(2\pi y/L)$ around $\theta = 0$ yields the condition
\begin{equation}
0 \simeq (C^2-1)L\tau + \int_0^L dy^\prime \sigma_{xy}
\end{equation}
and thus $\sigma_{xy} = \tau - 2\tau C^2 \sin^2(2\pi y/L) = \tau (1-2\delta\theta^2)$.
This determines $\sigma_{xy}$ and can be used in Eq. \eqref{eq:ThirdOrderDeltaEq} to find
\begin{equation}
0=\delta\theta^3-\frac{3 (\lambda - 1)}{2 (4\lambda-1)} \left(1-\frac{L_c^2}{L^2} \right)\delta\theta \,.
\end{equation}
Comparing with the normal form given in Sec. \ref{sec:4B3} we find that, for $L>L_c$, $\mu = 3(\lambda-1)/(2(4\lambda-1))$ is positive for all $\lambda < 1/4$ and $\lambda > 1$ and vanishes for $\lambda = 1$. 
With $\delta\theta=C\sin 2\pi y/L$ solving for $C$ at $y = L/2$ and expanding about $L \sim L_c$ to leading order yields the trivial solution $C=0$ and 
\begin{equation}
C \approx \pm \sqrt{\left(\frac{L}{L_c}-1\right)\left[\frac{3}{4+3/(\lambda-1)}\right]}\;,
\end{equation}
which has a singularity at $\lambda = 1/4$.
As remarked on in the general discussion in Sec. \ref{sec:4B3} the system displays a supercritical bifurcation for $\mu > 0$ but a subcritical bifurcation for $\mu < 0$. Thus the character of the bifurcation is determined by the sign of $\mu$ and in the present case this sign changes at the singularity $\lambda = 1/4$ and at $\lambda = 1$, where the system transitions between flow-tumbling ($\lambda < 1$) and flow-aligning ($\lambda>1$).
 
\bibliographystyle{vancouver}

\begin{thebibliography}{36}
	
\bibitem{Batchelor:1970}
G. K. Batchelor,
\href{https://doi.org/10.1017/S0022112070000745}{\emph{J. Fluid Mech.}, 1970, {\bf 41}, 545}.	

\bibitem{Duclos:2014}
G. Duclos, S. Garcia, H. G. Yevick and P. Silberzan,
\href{http://dx.doi.org/10.1039/c3sm52323c}{\emph{Soft Matter}, 2014, {\bf 10}, 2346}.

\bibitem{Garcia:2015}
S. Garcia, E. Hannezo, J. Elgeti, J. F. Joanny and P. Silberzan, N. S. Gov,
\href{http://dx.doi.org/10.1073/pnas.1510973112}{\emph{Proc. Natl. Acad. Sci. U.S.A.}, 2015, {\bf 112}, 15314}.

\bibitem{Duclos:2016}
G. Duclos, C. Erlenk\"amper, J. F. Joanny and P. Silberzan,
\href{http://dx.doi.org/10.1038/nphys3876}{\emph{Nat. Phys.}, 2016, {\bf 13}, 58}.	
	
\bibitem{Saw:2017}
T. B. Saw, A. Doostmohammadi, V. Nier, L. Kocgozlu, S. Thampi, Y. Toyama, P. Marcq, C. T. Lim, J. M. Yeomans and B. Ladoux,
\href{http://dx.doi.org/10.1038/nature21718}{\emph{Nature}, 2017, {\bf 544}, 212}.	
	
\bibitem{Kawaguchi:2017}
K. Kawaguchi, R. Kageyama and M. Sano,
\href{https://doi.org/10.1038/nature22321}{\emph{Nature}, 2017, {\bf 545}, 327}.	
	
\bibitem{Blanch-Mercader:2018}
C. Blanch-Mercader, V. Yashunsky, S. Garcia, G. Duclos, L. Giomi and P. Silberzan,
\href{https://doi.org/10.1103/PhysRevLett.120.208101}{\emph{Phys. Rev. Lett.}, 2018, {\bf 120}, 208101}.	
	
\bibitem{Duclos:2018}
G. Duclos, C. Blanch-Mercader, V. Yashunsky, G. Salbreux, J.-F. Joanny, J. Prost and P. Silberzan,
\href{http://dx.doi.org/10.1038/s41567-018-0099-7}{\emph{Nat. Phys.}, 2018, {\bf 14}, 7}.	
	
\bibitem{Doostmohammadi:2018}	
A.Doostmohammadi, J. Ign\'es-Mullol, J. M. Yeomans and F. Sagu\'es, 
\href{https://doi.org/10.1038/s41467-018-05666-8}{\emph{Nat. Commun.}, 2018, {\bf 9}, 3246}.	
	
\bibitem{Kemkemer:2000a}
R. Kemkemer, D. Kling, D. Kaufmann and H. Gruler,
\href{https://doi.org/10.1007/s101890050024}{\emph{Eur. Phys. J. E}, 2000, {\bf 1}, 215}. 

\bibitem{Kemkemer:2000b}
R. Kemkemer, V. Teichgräber, S. Schrank-Kaufmann, D. Kaufmann and H. Gruler,
\href{https://doi.org/10.1007/s101890070023}{\emph{Eur. Phys. J. E}, 2000, {\bf 3}, 101}.	

\bibitem{Giomi:2013}
L. Giomi, M. J. Bowick, X. Ma and M. C. Marchetti,
\href{http://dx.doi.org/10.1103/PhysRevLett.110.228101}{\emph{Phys. Rev. Lett.}, 2013, {\bf 110}, 228101}.

\bibitem{Giomi:2014}
L. Giomi, M. J. Bowick, P. Mishra, R. Sknepnek and M. C. Marchetti,
\href{http://dx.doi.org/10.1098/rsta.2013.0365}{\emph{Phil. Trans. R. Soc. A}, 2014, {\bf 372}, 20130365}.
	
\bibitem{Giomi:2015}
L. Giomi,
\href{https://doi.org/10.1103/PhysRevX.5.031003}{\emph{Phys. Rev. X}, 2015, {\bf 5}, 031003}.	
	
\bibitem{Henkes:2019}
S. Henkes, K. Kostanjevec, J. M. Collinson, R. Sknepnek and E. Bertin,
\href{https://arxiv.org/abs/1901.04763}{arXiv:1901.04763, 2019}.	
	
\bibitem{Voituriez:2005}
R. Voituriez, J.-F. Joanny and J. Prost,
\href{https://doi.org/10.1209/epl/i2004-10501-2}{\emph{Europhys. Lett.}, 2005, {\bf 70}, 404}.

\bibitem{Furthauer:2012a}
S. F\"urthauer, M. Strempel, S. W. Grill and F. J\"ulicher,
\href{ https://doi.org/10.1140/epje/i2012-12089-6}{\emph{Eur.Phys. J. E}, 2012, {\bf 35}, 89}.

\bibitem{Furthauer:2012b}
S. F\"urthauer, M. Strempel, S. W. Grill and F. J\"ulicher,
\href{https://doi.org/10.1103/PhysRevLett.110.048103}{\emph{Phys. Rev. Lett.}, 2012, {\bf 110}, 048103}.

\bibitem{Banerjee:2017}
D. Banerjee, A. Souslov, A. G. Abanov and V. Vitelli,
\href{https://doi.org/10.1038/s41467-017-01378-7}{\emph{Nat. Commun.}, 2017, {\bf 8}, 1573}. 

\bibitem{Avron:1995}
J. E. Avron, R. Seiler and P. G. Zograf,
\href{https://doi.org/10.1103/PhysRevLett.75.697}{\emph{Phys. Rev. Lett.}, 1995, {\bf 75}, 697}.

\bibitem{Pedley:1992}
T. J. Pedley and J. O. Kessler,
\href{http://dx.doi.org/10.1146/annurev.fl.24.010192.001525}{\emph{Annu. Rev. Fluid. Mech.}, 1992, {\bf 24}, 313}.

\bibitem{Simha:2002}
R. A. Simha and S. Ramaswamy,
\href{http://dx.doi.org/10.1103/PhysRevLett.89.058101}{\emph{Phys. Rev. Lett.,} 2002, {\bf 89}, 058101}.
	
\bibitem{Lau:2009}
A. W. C Lau and T. C. Lubensky,
\href{http://dx.doi.org/10.1103/PhysRevE.80.011917}{\emph{Phys. Rev. E}, 2009, {\bf 80}, 011917}.	
	
\bibitem{Wan:2011}	
L. Q. Wan, K. Ronaldson, M. Park, G. Taylor, Y. Zhang, J. M. Gimble and G. Vunjak-Novakovic,
\href{https://doi.org/10.1073/pnas.1103834108}{\emph{Proc. Natl. Acad. Sci}, 2011, {\bf 108}, 12295}.

\bibitem{Tee:2015}
Y. H. Tee, V. Thiagarajan, R. F. Hariadi, K. L. Anderson, C. Page, N. Volkmann, D. Hanein, S. Sivaramakrishnan, M. M. Kozlov and A. D. Bershadsky,
\href{https://doi.org/10.1038/ncb3137}{\emph{Nat. Cell Biol.}, 2015, {\bf 17}, 445}.

\bibitem{Schakenraad:2019}
K. Schakenraad, J. Ernst, W. Pomp, E. H. J. Danen, R. M. H. Merks, T. Schmidt and L. Giomi, 
\href{https://arxiv.org/abs/1905.09805}{arXiv:1905.09805, 2019}.
	
\bibitem{Worely:2015}	
K. E. Worley, D. Shieh and L. Q. Wan,
\href{https://doi.org/10.1039/c5ib00073d}{\emph{Integr. Biol.}, 2015, {\bf 7}, 580}.

\bibitem{Sanchez:2012}
T. Sanchez, D. N. Chen, S. J. DeCamp, M. Heymann and Z. Dogic,
\href{http://dx.doi.org/10.1038/nature11591}{\emph{Nature}, 2012, {\bf 491}, 431}.

\bibitem{Keber:2014}
F. C. Keber, E. Loiseau, T. Sanchez, S. J. DeCamp, L. Giomi, M. J. Bowick, M. C. Marchetti, Z. Dogic and A. R. Bausch,
\href{http://dx.doi.org/10.1126/science.1254784}{\emph{Science}, 2014, {\bf 345}, 1135}.

\bibitem{Guillamat:2016}
P. Guillamat, J. Ign\'es-Mullol and Francesc Sagu\'es,
\href{https://doi.org/10.1073/pnas.1600339113}{\emph{Proc. Natl. Acad. Sci. U.S.A}., 2016, {\bf 113}, 5498}.

\bibitem{Guillamat:2017}
P. Guillamat, J. Ign\'es-Mullol and Francesc Sagu\'es,
\href{https://doi.org/10.1038/s41467-017-00617-1}{\emph{Nat. Commun}., 2017, {\bf 8}, 564}.

\bibitem{Zhang:2018}
R. Zhang, N. Kumar, J. L. Ross, M. L. Gardel and J. J. de Pablo,
\href{https://doi.org/10.1073/pnas.1713832115}{\emph{Proc. Natl. Acad. Sci. U.S.A.}, 2018, {\bf115}, E124}.
	
\bibitem{Vromans:2016}
A. J. Vromans and L. Giomi,
\href{https://doi.org/10.1039/C6SM01146B}{\emph{Soft Matter}, 2016, {\bf 12}, 6490}.

\bibitem{Maitra:2019}
A. Maitra and M. Lenz,
\href{https://doi.org/10.1038/s41467-019-08914-7}{\emph{Nat. Commun.}, 2019, {\bf 10}, 920}.

\bibitem{Marenduzzo:2007}
D. Marenduzzo, E. Orlandini, M. E. Cates and J. M. Yeomans,
\href{https://doi.org/10.1103/PhysRevE.76.031921}{\emph{Phys. Rev. E}, 2007, {\bf 76}, 031921}.

\bibitem{Giomi:2008}
L. Giomi, M. C. Marchetti and T. B. Liverpool,
\href{https://doi.org/10.1103/PhysRevLett.101.198101}{\emph{Phys. Rev. Lett.}, 2008, {\bf 101}, 198101}.

\bibitem{Edwards:2009}
S. A. Edwards and J. M. Yeomans,
\href{https://doi.org/10.1209/0295-5075/85/18008}{\emph{Europhys. Lett.}, 2009, {\bf 85}, 18008}.

\bibitem{Hardouin:2019}
J. Hardo\"uin, R. Hughes, A. Doostmohammadi, J. Laurent, T. Lopez-Leon, J. M. Yeomans, J. Ign\'es-Mullol and F. Sagu\'es,
\href{https://arxiv.org/abs/1903.01787}{arXiv:1903.01787, 2019}.

\bibitem{Freedericksz:1927}
V. Fr\'eedericksz and A. Repiewa,
\href{https://doi.org/10.1007/BF01397711}{\emph{Z. Physik}, 1927, {\bf 42}, 532.} 

\bibitem{DeGennes:1993}
P. G. de Gennes and J. Prost,
{\em The Physics of Liquid Crystals}, Oxford University Press, Oxford, 1993.

\bibitem{Golubitsky:1985}
M. Golubitsky and D. G. Schaeffer,
{\em Singularities and Groups in Bifurcation Theory}, Vol. I, Springer-Verlag, New York, 1985.

\bibitem{Kitanov:2013}
P. M. Kitanov, W. F. Langford and A. R. Willms,
\emph{Dynam. Cont. Dis. Ser. A}, 2013, {\bf 20}, 197.

\bibitem{Blow:2014}
M. L. Blow, S. P. Thampi and J. M. Yeomans,
\href{https://doi.org/10.1103/PhysRevLett.113.248303}{\emph{Phys. Rev. Lett.,} 2014, {\bf 113}, 248303}.

\bibitem{Doostmohammadi:2016}
A. Doostmohammadi, S. P. Thampi and J. M. Yeomans,
\href{https://doi.org/10.1103/PhysRevLett.117.048102}{\emph{Phys. Rev. Lett.}, 2016, {\bf 117}, 048102}.

\end{thebibliography}

\end{document}